\documentclass[sigconf, screen]{acmart}

\AtBeginDocument{%
  \providecommand\BibTeX{{%
    \normalfont B\kern-0.5em{\scshape i\kern-0.25em b}\kern-0.8em\TeX}}}

\copyrightyear{2023} 
\acmYear{2023} 
\setcopyright{acmlicensed}\acmConference[AutomotiveUI '23]{15th International Conference on Automotive User Interfaces and Interactive Vehicular Applications}{September 18--22, 2023}{Ingolstadt, Germany}
\acmBooktitle{15th International Conference on Automotive User Interfaces and Interactive Vehicular Applications (AutomotiveUI '23), September 18--22, 2023, Ingolstadt, Germany}
\acmPrice{15.00}
\acmDOI{10.1145/3580585.3607152}
\acmISBN{979-8-4007-0105-4/23/09}

\begin{document}

\title[Assessing AR Selection Techniques for Passengers: A Real-World User Study]
{Assessing Augmented Reality Selection Techniques for Passengers in Moving Vehicles: A Real-World User Study}



\author{Robin Connor Schramm}
\email{robin.schramm@mercedes-benz.com}
\orcid{0000-0002-4775-4219}
\affiliation{
  \institution{Mercedes-Benz Tech Motion GmbH}
  \streetaddress{Gutenbergstraße 19}
  \city{Leinfelden-Echterdingen}
  \country{Germany}
  \postcode{70771}
}

\author{Markus Sasalovici}
\email{markus.sasalovici@mercedes-benz.com}
\orcid{0000-0001-9883-2398}
\affiliation{
  \institution{Mercedes-Benz Tech Motion GmbH}
  \streetaddress{Gutenbergstraße 19}
  \city{Leinfelden-Echterdingen}
  \country{Germany}
  \postcode{70771}
}

\author{Axel Hildebrand}
\email{axel.hildebrand@mercedes-benz.com}
\orcid{0009-0008-3038-7775}
\affiliation{
  \institution{Mercedes-Benz Tech Motion GmbH}
  \streetaddress{Gutenbergstraße 19}
  \city{Leinfelden-Echterdingen}
  \country{Germany}
  \postcode{70771}
}

\author{Ulrich Schwanecke}
\email{ulrich.schwanecke@hs-rm.de}
\orcid{0000-0002-0093-3922}
\affiliation{
  \institution{Hochschule RheinMain}
  \streetaddress{Unter den Eichen 5}
  \city{Wiesbaden}
  \country{Germany}
  \postcode{65195}
}

\renewcommand{\shortauthors}{Schramm et al.}

\begin{abstract}
  Nowadays, cars offer many possibilities to explore the world around you by providing location-based information displayed on a 2D-Map. However, this information is often only available to front-seat passengers while being restricted to in-car displays. To propose a more natural way of interacting with the environment, we implemented an augmented reality head-mounted display to overlay points of interest onto the real world. We aim to compare multiple selection techniques for digital objects located outside a moving car by investigating head gaze with dwell time, head gaze with hardware button, eye gaze with hardware button, and hand pointing with gesture confirmation. Our study was conducted in a moving car under real-world conditions (N=22), with significant results indicating that hand pointing usage led to slower and less precise content selection while eye gaze was preferred by participants and performed on par with the other techniques.
\end{abstract}



\begin{teaserfigure}
  \includegraphics[width=\textwidth]{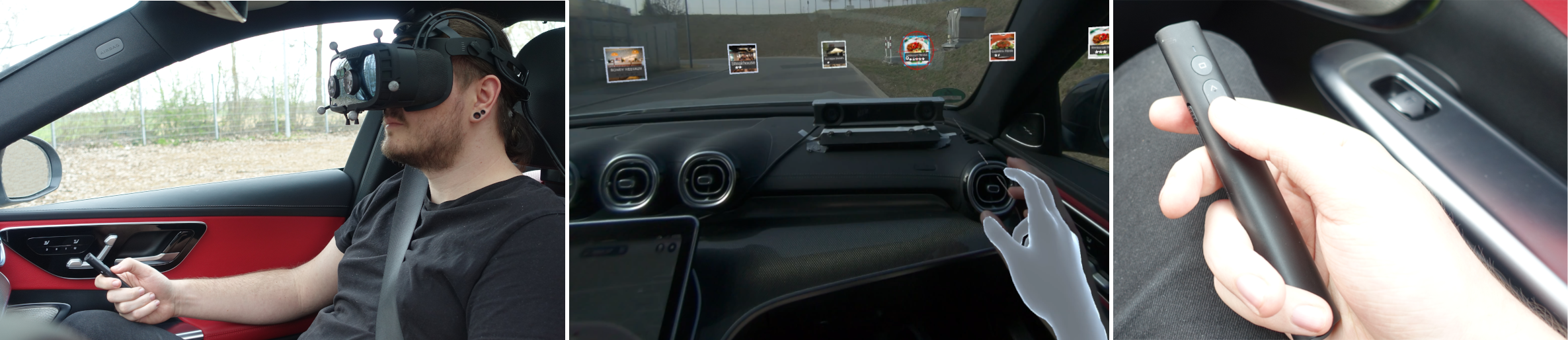}
  \Description{The user study setup divided into three images. Shown in the left image, a person sits in the passenger seat of a car wearing a Varjo XR-3 HMD on their head and a wireless presenter in their hand. In the middle image, the view through the HMD is shown. Outside the car, digital objects are overlaid onto the real world. In the right image, the wireless presenter can be seen in more detail, held in the right hand.}
  \caption{User study setup. From left to right: in-car AR HMD setup, AR point-of-view during the study, mobile presenter for hardware selection.}
  \Description{}
  \label{fig:teaser}
\end{teaserfigure}


\maketitle

\section{Introduction}
The automotive sector offers many opportunities for augmented reality (AR) use-cases such as navigation, information, and entertainment, e.g. in the form of points of interest (POIs). Consequently, the adoption of AR technology in vehicles has been growing rapidly, with major car manufacturers like Audi, BMW, and Mercedes-Benz already incorporating Head-Up Displays (HUDs) and video AR features to display navigation data and other information. Researchers are also examining usage of windshield AR displays which would cover the whole front view \cite{haeuslschmid2016design}. Over time, AR hardware is becoming lighter while offering a bigger field-of-view (FoV). With those advances, both Head-Mounted Displays (HMDs) and AR glasses are becoming increasingly suitable for in-car use and should be taken into consideration by automotive Human-Computer Interaction (HCI) researchers \cite{riegler2021augmented}. AR systems can provide passengers with context specific, real-time, information about their surroundings, such as traffic conditions, POIs, and hazards overlaid on their view of the real world. Data like this puts digital content into a real-world context and makes it potentially easier to understand information at a glance \cite{haeuslschmid2016design}. Furthermore, the emergence of autonomous vehicles means that passengers will have a considerable amount of free time while in transit \cite{mcgill2020challenges} to work \cite{Mathis2021work} or to play games \cite{Togwell2022gaming}.

Additionally, HMDs offer a variety of built-in features for interacting with digital objects, such as head-tracking, hand-tracking, and eye-tracking sensors. Leveraging these natural interaction methods can offer several advantages over traditional touch-based methods, such as increased simplicity and reduced distraction, e.g. while driving \cite{gomaa2020studying}. However, in order for users to effectively interact with their surroundings, they require precise methods for selecting and manipulating objects and UI elements, which becomes especially critical in dynamic environments such as a moving vehicle.

In this work, we evaluate selection techniques for interacting with digital objects outside the car while in motion using HMD-based AR. We build on related work by adapting existing AR selection techniques to an in-car environment. Based off widely researched interaction techniques \cite{kyto2018pinpointing}, we investigate the efficacy of the following four techniques: head pointing with dwell time, head pointing with a hardware button, eye gaze with a hardware button, and hand pointing with gesture confirmation. We provide a comprehensive analysis of the results and discuss their implications for researchers and designers who aim to integrate AR-HMDs into future automotive concepts and applications. Thus, the main contributions of this paper are as follows:
\begin{itemize}
    \item A comparison of four selection techniques for use with an AR-HMD in a moving vehicle, tested under real-world conditions.
    \item Detailed analysis of each interaction technique, by portraying individual advantages or disadvantages concerning the dimensions speed, error rate, workload, and usability.
    \item Evaluation of selection techniques and an HMD-based AR system for usage in moving vehicles by means of a semi-structured interview.
\end{itemize}

\section{Related work}
\label{chapter:related}
Selection is an essential task for the interaction between a user and virtual elements across mixed reality (MR) systems \cite{blattgerste2018advantages,Doerner2022}. While selecting objects in MR has been extensively studied \cite{nizam2018review, hertel2021taxonomy,kyto2018pinpointing, zhou2008trends, blattgerste2018advantages}, the majority of research in this field is focused solely on usage within stationary environments. Nevertheless, insights from this research can still be applied to HCI research in dynamic environments. We focus specifically on remote interaction techniques considering the use-case of POIs outside the vehicle. For example, Blattgerste et al. \cite{blattgerste2018advantages} conducted a study comparing the performance of head gaze and eye gaze during the aiming phase. They found that eye gaze outperformed head gaze in terms of speed, task load, required head movement and user preference. They also discovered that the advantages of eye gaze increased with larger FoV sizes.
Tanriverdi and Jacob \cite{tanriverdi2000interacting} found that eye gaze-based interaction was faster than hand pointing, even with hardware limitations of the time. On the contrary, Cournia et al. \cite{cournia2003gaze} found that gaze-based techniques were slower than hand pointing for distant objects. Luro and Sundstedt~\cite{luro2019comparative} found similar results between eye gaze and a hand controller regarding usability, while eye gaze induced a lower cognitive load in participants. Kyoto et al.~\cite{kyto2018pinpointing} extensively compared head pointing and eye gaze in combination with various refinement techniques. They found eye gaze to be faster and head pointing to be more accurate. Furthermore, participants preferred hardware-based inputs to confirm selections over gestures. Hansen et al. \cite{hansen2018fitts} compared eye gaze, head gaze, and mouse input combined with dwell time and click for confirmation in regard to Fitt's Law. They found eye gaze to be less accurate and having a lower throughput, while head pointing was more physically demanding. 
Techniques that utilize dwell time can suffer from the Midas touch problem \cite{jacob1990you,Doerner2022}, where involuntary selections can occur due to simply looking around, e.g. via head gaze or eye gaze.

There are also existing works that studied the selection of objects from the inside of a vehicle. However, previous works mostly study interaction with objects inside the car \cite{aftab2020point}, during short driving rounds \cite{gomaa2020studying}, or don't use AR. For example, Rümelin et al. \cite{rumelin2013free} and Fujimura et al. \cite{fujimura2013driver} investigated hand pointing for interaction with distant objects while in the vehicle, but did not use AR HMDs. Aftab et al. \cite{aftab2021multimodal} explored multimodal interaction of head, eye, and finger direction for drivers referencing outside-vehicle objects. Gomaa et al.~\cite{gomaa2020studying} also adopted a multimodal approach to use eye gaze and hand pointing to reference objects outside the car. While hand-based interaction has been commonly studied inside vehicles, McGill et al. \cite{mcgill2020challenges} pointed out that physical constraints of the in-car environment, such as available space and motion restraints, may impair the effectiveness of such techniques.

There are also challenges regarding HMD use for vehicles in transit. McGill et al. \cite{mcgill2020challenges} examined challenges of HMD use for passengers in cars and other transportation systems. While motion sickness, safety and technical obstacles are examples of challenges to overcome, they argued that it is justified to further explore the use of MR headsets in cars. Their reasoning mainly consists of a potentially improved passenger experience in terms of productivity, entertainment and isolation. Furthermore, Riegler et al. \cite{riegler2020agenda} proposed a research agenda for MR in automated vehicles. One of their overarching points was to investigate the use of HMDs to create useful in-car work and entertainment experiences for passengers.

Usually, in-vehicle HCI studies are tested using some form of simulators such as VR simulators \cite{colley2022swivr}, CAVE systems or professional driving simulators. Virtual driving simulators can have value, but can not replace real-world studies \cite{petterson2019virtually}. A driving simulator may for example influence the participant's behavior as it is a less stressful environment compared with a moving vehicle in traffic \cite{gomaa2020studying}. Driving simulators are also constricted regarding realism, as they can not match the fidelity of the real world \cite{kun2018hcivehicles}.

\section{Study}\label{chapter:study}
In this section, we describe the user study we conducted to compare different selection methods for AR in moving vehicles. We decided to investigate four input methods within our study, which to the best of our knowledge have not yet been fully explored in the context of a moving vehicle. These methods are head gaze with dwell time (HeadDwell), head gaze with hardware button (HeadHardware), eye gaze with hardware button (EyeHardware), and hand pointing with gesture confirmation via airtap (HandPoint). This design decision to only investigate four techniques was motivated by the need to maintain a sufficiently short duration for the overall experiment, in order to mitigate the risk of terminations due to the risk of motion sickness.

\subsection{Research Questions}
Due to the four distinct methods, we chose a within-subject design. We evaluated four selection methods for digital objects in AR with the following research questions:
\begin{itemize}
    \item \textbf{Q1:} \textit{Which selection method offers the highest selection speed and lowest error rate?}
    \item \textbf{Q2:} \textit{Does the selection method have an impact on perceived workload and usability?}
    \item \textbf{Q3:} \textit{Which selection methods are the most and least preferred by participants?}
\end{itemize}

\subsection{Participants and Apparatus} \label{chapter:participants_and_apparatus}
We recruited 23 participants for the study, but had to exclude one participant's data due to terminating the experiment early because of motion sickness. This resulted in a number of 22 participants (18 male, 4 female) with a mean age of 36.3 years ($SD=10.6$). All participants were employees of an automotive software consulting company. Ten participants required prescription glasses and seven of them didn't use them during the study because the glasses didn't fit comfortably inside the HMD. Participants were also asked to report their experience with immersive technologies, such as VR or AR. 15 participants had at least some experience in developing immersive technologies and therefore had used those technologies in the past, 5 people had experience using immersive technologies and 3 participants had never used any VR or AR headset. The mean study duration per participant was about 80 minutes.

\begin{figure}
    \centering
    \includegraphics[width=0.35\textwidth]{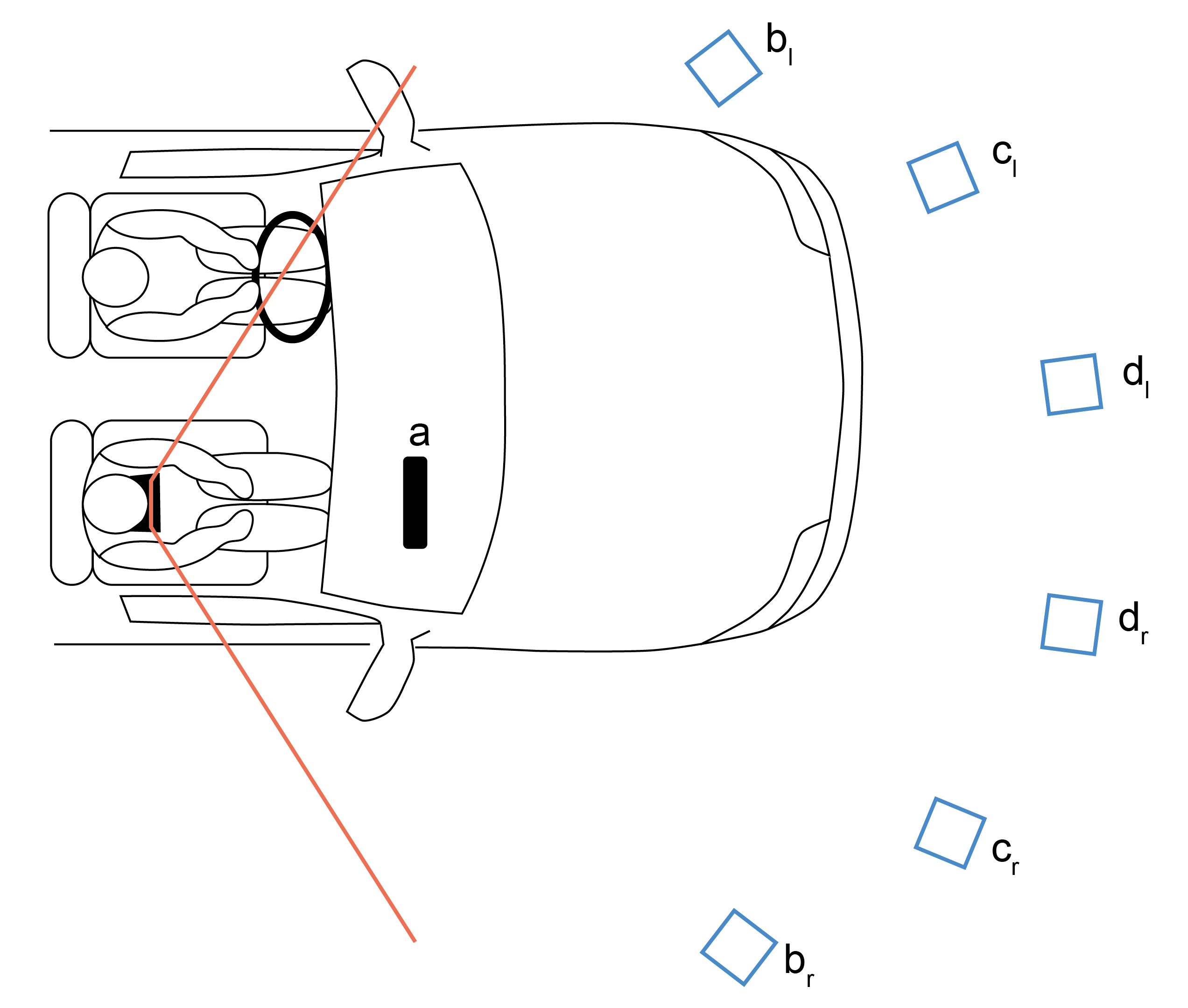}
    \Description{A drawn schematic, showing a top-down view of a car with a driver and a passenger inside. The passenger wears an HMD. The field-of-view of the HMD is visualized with red lines. In front of the car, six labeled squares represeting points of interest are placed as described in the paper. All squares are within the HMDs field-of-view.}
    \caption{A schematic of the set-up of our user study. The participant sat on the passenger seat. The red lines visualize the FoV of the Varjo XR-3. a) shows the position of the optical tracking system, b) to d) visualize the position and rotation of the POIs at the representative angles. Distances are not perfectly to scale.}
    \label{fig:STUDY_SCHEMATIC}
\end{figure}

The set-up of our study is shown in Figure \ref{fig:STUDY_SCHEMATIC}. The participants were seated in the front passenger seat of a premium midsize estate. In order to carry out the study under realistic and controllable conditions, we chose a private traffic-calmed environment. The environment resembled an urban area with some traffic in form of cars, bikes, as well as pedestrians crossing streets. The car's speed limiter has been set to the maximum allowed speed in the study environment of 30 km/h to support uniform driving conditions.

For the AR-hardware, we chose the Varjo XR-3 video see-through HMD~\cite{VarjoXR3} based on its state-of-the-art resolution, 115° FoV, 90Hz refresh rate, video see-through latency of < 20ms as well as for its features, namely built-in eye tracking, and hand tracking via an integrated Ultraleap Gemini~\cite{UltraleapGemini}. Six degrees of freedom (6-DoF) HMD tracking was achieved by using optical tracking and an additional car-fixed Inertial Measurement Unit. The system was connected to a Varjo XR-3 compatible desktop PC running the required Varjo software, the study software in Unity, and the tracking software required for 6-DoF tracking. The selection techniques were implemented based on Microsoft's Mixed Reality Toolkit (MRTK) Version 2.8.3~\cite{MRTK}. MRTK-based interactions proved themselves within a wide range of contexts and industry projects and could therefore provide a reliable implementation of the methods. The dwell time for the HeadDwell condition was set to one second, based on the findings of Riegler et al.~\cite{riegler2020gaze}. Dwelling progress did not reset to zero immediatly after losing focus, instead it decayed over the span of two seconds in case car movements influenced the participant's head position. For selections that required hardware input, a simple three-button portable Bluetooth presenter was used, as seen in Figure~\ref{fig:teaser}. After hovering a POI either via head gaze or eye gaze depending on the condition tested, the user simply needed to press one of the three buttons to trigger the selection. For the HandPoint condition, we used MRTK's integrated LeapMotion support with the default air tap gesture for remote interaction. The system allowed Users to use both hands in any way they wanted.

\subsection{Stimuli and Procedure}
At the beginning of each study session, participants signed a privacy and information sharing consent agreement and completed a demographic questionnaire. This was followed by an introduction to the experiments procedure and the task. Participants were not informed about the selection methods before the study in order to reduce possible influence on the quantitative data through comparison or the qualitative feedback through preference bias.

Participants wore a Varjo-XR 3 in video see-through mode as described in Chapter \ref{chapter:participants_and_apparatus} and were positioned on the front passenger seat. In each trial, six digital UI elements were shown outside the car in front of the participant, as shown in Figure \ref{fig:STUDY_SCHEMATIC}. The UI elements resembled restaurant POIs filled with randomized fake data. Each POI consisted of a 2D square sized about 4.14° in angular distance, an image with a white border and the restaurant name, which can be seen in Figure \ref{fig:teaser}. POIs were rotated to always face the camera. The POI's center points were placed on a half circle with a target angle of 15° shifted to the left and right, which resulted in angles of $\pm7.5$°, $\pm22.5$°, and $\pm37.5$°. This results in uniform 10.86° sized gaps between each pair of POIs. 

The POIs were in car-fixed space and thus followed the translation and rotation of the car. Between each input method, the car was halted and participants were introduced to the next method. This took part in form of an explanation and a training session, where participants could test the upcoming method on six unmarked POIs. During this training time, the system didn't record any data. After the participant had successfully selected all six POIs, the trial started automatically and the ride began. There were four trials per participant, one for each of the four selection methods. In order to prevent learning effects, the order of trials was counterbalanced using Latin square. 
Before each trial began, participants were instructed to think aloud in order to provide live feedback on situations. We used these comments to enrich the post-study interview. All steps were recorded on video for later analysis. Each trial consisted of 70 rounds, each round lasting five seconds with a one-second pause in between. Six POIs appeared in each round, one of which was marked randomly with a red crosshair. The goal of each round was to select the marked POI without selecting an unmarked POI. After each round, all POIs, whether marked or not, disappeared for a second, then a new set of six POIs appeared with one POI marked. The route was roughly timed to match the speed of the 70 rounds, but was dependent on traffic. The car was stopped after completing the course, regardless whether the run was finished. In that case, participants proceeded with the last few rounds while standing still. After all rounds were completed, a message with the text "run finished" appeared and the car was stopped, with participants being allowed to remove the HMD. Following each trial, the participants filled out the questionnaires described in chapter \ref{chapter:measures} and were asked what they noticed positively and negatively, as well as how much they perceived their surroundings.

\subsection{Measures}\label{chapter:measures}
We chose the System Usability Scale (SUS) \cite{Brooke.1996} and the NASA Task Load Index (TLX) \cite{hart1988development} as quantitative metrics. The SUS provides an overall usability score by a formula, which indicates the system's overall usability. Scores range from 0 to 100, with higher scores being better. Bangor et al. \cite{bangor2009sus} linked SUS scores with adjectives such as good or excellent, which will be used as a basis for comparing and discussing the results. For the NASA TLX, the Raw TLX (RTLX) variant \cite{byers1989traditional} was used for simplicity. With RTLX, each subscale is given a score between 0 and 100, while the total score simply consists of the sum of these numbers and thus ranges between 0 and 600. A higher score indicates higher perceived workload by the user. It is also possible to analyze the individual subscales in addition to the overall score \cite{hart2006nasa}.

For performance evaluation, the system recorded each hover and selection with respective timestamps, round, selection technique, and information on whether the POI hovered or selected was the marked one. For the techniques using a hardware button, each button click was also documented with a timestamp. 
Based on this data, the elapsed time between each hover and subsequent selection was also calculated, which gives insight into the speed of buttons presses and air tap gestures after hovering. This metric will be called DeltaSelectHover going further. Additionally, an error rate was calculated based on how many hovers and selections were correct or incorrect.
Furthermore, we collected qualitative data in the form of comments made by participants during the study and through semi-structured interviews. The interview questions after each round asked for positive and negative feedback on the technique used, how the environment was perceived, and technique-specific questions such as feedback on the dwell time. At the end of the study, participants had to name their favorite and least favorite technique, as well as provide feedback on the HMD and their opinion on AR in moving vehicles.

\section{Results}
\label{chapter:results}
We used a one-way ANOVA with the selection method as independent variable and RTLX scores, SUS scores and selection times as dependent variables. When the assumptions of normality or the assumption of homogeneity  of variances was violated, we used the non-parametric Kruskal-Wallis H test. Post-hoc tests were then conducted using Dwass-Steel-Critchlow-Fligner (DSCF) pairwise comparisons. Only significant results assuming a 5\% significance level or otherwise notable results are reported.

\subsection{Time}
The selection technique had a significant effect on the mean selection time ($\chi^2(3) = 1141, p < 0.001$) as well as the elapsed time between a hover and the subsequent selection ($\chi^2(3) = 2281, p < 0.001$). The mean, median, and standard deviation for each condition for selection time and the time between hover and selection is shown in Table \ref{tab:TIME_DESCRIPTIVES}. For select time, DSCF pairwise comparisons found that the mean value was significantly different for all groups with $p < 0.001$ for each pair. Similarly, DSCF pairwise comparisons found significant differences between all pairs with $p < 0.001$, except for the pair HeadHardware and EyeHardware ($W = 2.02, p = 0.480$). This shows, that the time between a hover and the subsequent selection via hardware button is not significantly influenced by the technique to hover the element. The techniques using a hardware button were overall faster regarding both the mean time participants needed for one selection and the delta between a hover and the following selection. Notably, selections via EyeHardware were fastest with a mean time of $1.82s$, followed by HeadHardware with $1.95s$. HandPoint was the slowest technique with a mean time of $2.76s$. Notably, the mean delta time of HeadDwell was found to be 0.993s, slightly below the one-second threshold specified in Unity. This is likely due to Unity's Time.deltaTime, which can vary depending on the framerate \cite{UnityTime}.

\begin{table}[h]
    \caption{Mean, median and standard deviation for the elapsed time per round in seconds (SelectTime) and the time between a hover and the subsequent selection in seconds (Hover-Select) for each method. The lowest and therefore best values are highlighted in bold.}
    \label{tab:TIME_DESCRIPTIVES}
    \begin{center}
    \begin{tabular}{llrr}
        \toprule
                            & \textbf{Condition}    & \textbf{SelectTime} & \textbf{Hover-Select} \\
        \midrule
        Mean                & HeadDwell    & 2.45       & 0.993            \\
                            & HandPoint    & 2.76       & 0.930            \\
                            & HeadHardware & 1.95       & \textbf{0.494}            \\ \medskip
                            & EyeHardware  & \textbf{1.82}       & \textbf{0.494}            \\ 
        Median              & HeadDwell    & 2.30       & 0.911            \\
                            & HandPoint    & 2.56       & 0.733            \\
                            & HeadHardware & 1.78       & \textbf{0.467}            \\ \medskip
                            & EyeHardware  & \textbf{1.53}       & 0.523            \\
        Standard  & HeadDwell    & 0.676      & 0.284            \\
        Deviation                    & HandPoint    & 0.975      & 0.646            \\
                            & HeadHardware & 0.741      & 0.249            \\
                            & EyeHardware  & 0.913      & 0.312            \\
        \bottomrule
    \end{tabular}
    \end{center}
\end{table}

\subsection{Error Rate}
Table \ref{tab:SELECTIONS} shows the respective correct, wrong and missing selections for each method. The maximum possible amount of correct selections was 1540. Notably, while using HandPointing, participants didn't select 30\% of the marked targets. This could stem from multiple factors, which will be discussed in chapter \ref{chapter:discussion}. Furthermore, with HeadDwell, participants selected 96.56\% of marked POIs with only 11 selections on unmarked POIs, even though HeadDwell was the only tested technique with implicit selection via dwell time.

\begin{table}[h]
    \renewcommand{\arraystretch}{1.3}
    \caption{Frequencies of POI selections grouped by correct selections (Marked), incorrect selections (Unmarked) number of rounds without any selections (No Selection). There were 1540 rounds in total per condition. The best values are highlighted in bold.}
    \begin{tabular}{p{1.9cm}ccl}
    \toprule
    \textbf{Condition} & \textbf{Marked} & \textbf{Unmarked} & \textbf{No Selection} \\
    \midrule
    HeadDwell          & \textbf{1487 (96.56\%)}  & 11 (0.71\%)    & \textbf{42 \ \ (2.73\%)}       \\
    HandPoint          & 1076 (69.87\%)  & \textbf{2 \ (0.13\%)}   & 462 (30.0\%)            \\
    HeadHardware       & 1458 (94.68\%)  & 5 \ (0.32\%)   & 77 \ \ (5.0\%)          \\
    EyeHardware        & 1376 (89.35\%)  & 44 (2.86\%)    & 120 (7.79\%)          \\
    \bottomrule
    \end{tabular}
    \label{tab:SELECTIONS}
\end{table}

\subsection{Perceived Workload} \label{chapter:Workload}
\begin{figure}
    \centering
    \includegraphics[width=0.4\textwidth]{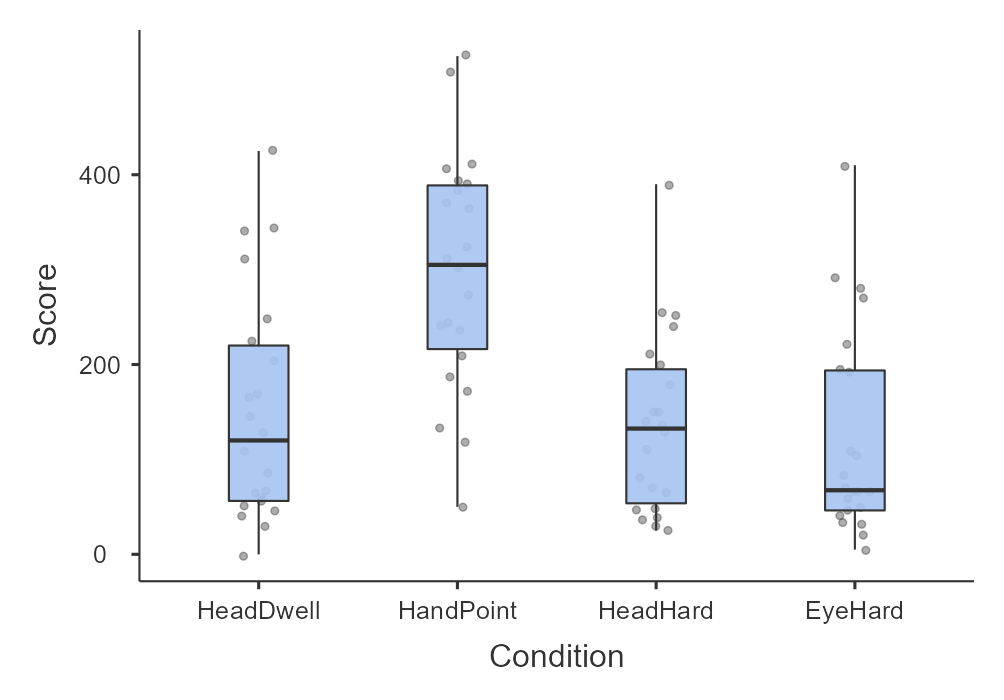}
    \Description{Boxplots for TLX values. All values are roughly under 200, except HandPoint, which hovers between 200 and 380 and thus stands out to the rest.}
    \captionof{figure}{The mean values for the RTLX total score grouped by condition. Lower is better.}
    \label{fig:BOXPLOT_TLX_SCORE}
\end{figure}

The analysis of the RTLX questionnaire showed significant differences between techniques for perceived workload regarding all six subscales and the total workload score. The results of the non-parametric Kruskal-Wallis H test are shown in Table \ref{tab:anova_rtlx}, mean values for each subscale grouped by selection technique are shown in Table \ref{tab:TLX_MEANS}. Total RTLX scores for each method are visualized in Figure \ref{fig:BOXPLOT_TLX_SCORE}. The median for EyeHardware is notably lower than the mean ($M = 123, Mdn = 67.5$). This could stem from minor problems a few select people had with the eye-tracking system, thus resulting in a higher mean and lower median value. This is also reflected most notably by the subscales performance ($M = 4.32, Mdn = 2.50$) and frustration ($M = 3.91, Mdn = 2.50$).

DSCF pairwise comparisons were conducted for each pairing of methods for each subscale and the score. Regarding mental workload, only HandPoint and EyeHardware had significant differences ($W = -4.022, p = 0.023$). EyeHardware was mentally the least demanding while HandPoint was mentally the most demanding, which is also reflected in statements in the qualitative interview.
Physical workload showed significant differences between HandPoint and HeadHardware ($W = -3.811, p = 0.036$) as well as between HandPoint and EyeHardware ($W = -5.594, p < 0.001$). This can be explained by the low physical effort the hardware button required. Selection times for techniques using the button were the fastest, thus participants only needed to move their head or eyes for a short time in the direction of the marked POI and could then move back into a resting position. In contrast, the average and median speed for HandPoint were the lowest, meaning that participants had to hover their arms longer before moving them back into a resting position, resulting in higher physical workload.

Performance, effort, frustration, temporal demand and the final score all had significant differences between HandPoint paired with each other technique. Pairings between HeadDwell, HeadHardware and EyeHardware showed no significant differences for those subscales. This difference between the HandPointing technique and the other techniques is also visible in mean scores.

\begin{table}
    \centering
        \caption{Means of RTLX scores per sub-scale grouped by selection methods. Lower is better.}
        \label{tab:TLX_MEANS}
        \begin{tabular}{p{1.45cm}cccc}
        \toprule
        Sub-scale   & HeadDwell & HandPoint & HeadHard & EyeHard \\
        \midrule
        Mental      & 5.05      & 7.64      & 4.68     & 3.64    \\
        Physical    & 6.50      & 9.95      & 5.50     & 3.36    \\
        Temporal    & 5.64      & 10.7      & 4.14     & 4.55    \\
        Performance & 2.50      & 10.0      & 3.55     & 4.32    \\
        Effort      & 5.68      & 10.7      & 5.91     & 4.82    \\
        Frustration & 4.77      & 10.6      & 3.32     & 3.91    \\
        \midrule
        Score       & 151       & 298       & 135      & 123     \\
        \bottomrule
        \end{tabular}
\end{table}

\begin{table}
    \centering
    \caption{Results of the Kruskal-Wallis H test for each RTLX subscale.}
    \label{tab:anova_rtlx}
    \begin{tabular}{lcccc}
        \toprule
        Sub-scale & $X^2$ & df & p & $\epsilon^2$ \\
        \midrule
        Score&22.30&3&<.001&0.2563\\
        Mental&8.21&3&0.042&0.0943\\
        Physical&15.54&3&0.001&0.1787\\
        Temporal&21.03&3& <.001&0.2417\\
        Performance&29.00&3&<.001&0.3334\\
        Effort&14.18&3&0.003&0.1630\\
        Frustration&19.33&3&<.001&0.2222\\
        \bottomrule
    \end{tabular}
\end{table}
     
\begin{table}
    \centering
    \caption{The mean, median and standard deviation values for the SUS score, grouped by condition. Higher is better.}
    \label{tab:descriptives_SUS}
    \begin{tabular}{lccc}
        \toprule
        \textbf{Condition} & \textbf{Mean} & \textbf{Median} & \textbf{SD} \\
        \midrule
        HeadDwell          & 83.8          & 90.0            & 15.0        \\
        HandPoint          & 58.0          & 53.8            & 23.1        \\
        HeadHard           & 87.6          & 90.0            & 10.8        \\
        EyeHard            & 85.6          & 88.8            & 17.8        \\
        \bottomrule
    \end{tabular}
\end{table}

\subsection{Usability} \label{chapter:Usability}
\begin{figure}
    \centering
    \includegraphics[width=0.4\textwidth]{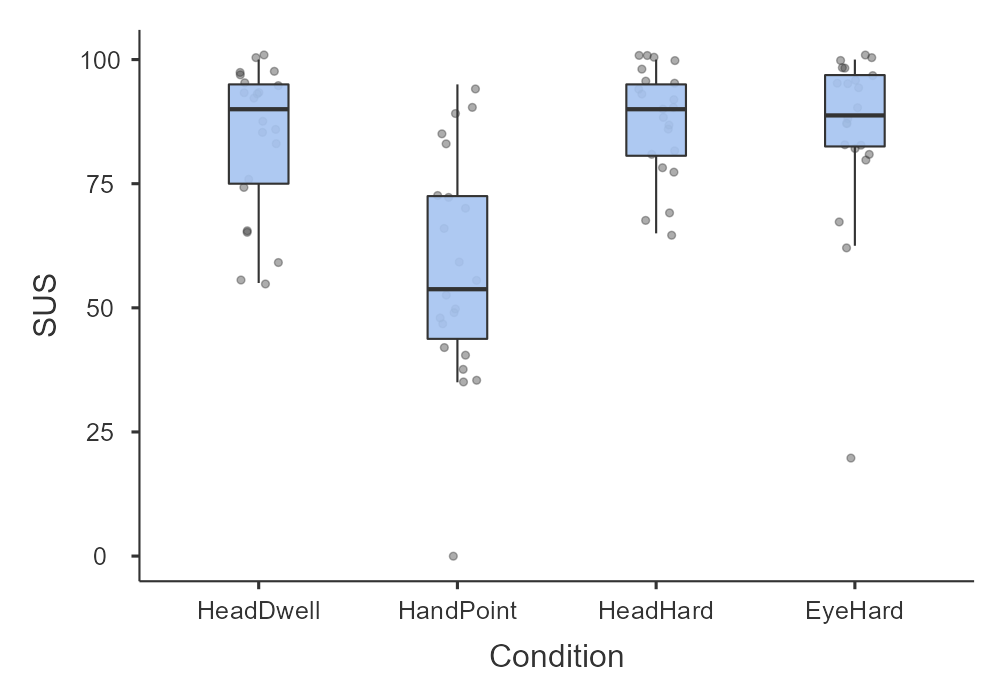}
    \Description{Boxplots for SUS values. All values are roughly between 75 and 90 and shown at the top, except HandPoint, which hovers between 45 and 75 and thus is placed significantly lower.}
    \caption{The mean values for the SUS score grouped by condition. Higher is better.}
    \label{fig:BOXPLOT_SUS_SCORE}
\end{figure}

There was a statistically significant difference between  methods regarding usability scores ($\chi^2(3) = 25.3, p < 0.001$). Figure \ref{fig:BOXPLOT_SUS_SCORE} and Table \ref{tab:descriptives_SUS} show the data for SUS scores. DSCF pairwise comparisons showed significant differences between HandPoint and HeadDwell ($W = -5.299, p = 0.001$), HandPoint and HeadHard ($W = 5.868, p < 0.001$) and HandPoint and EyeHard ($W = 5.953, p < 0.001$), which is also visible in the mean SUS scores.

\subsection{User Preferences} \label{chap:user_preference}
\begin{figure}
    \centering
    \includegraphics[width=0.4\textwidth]{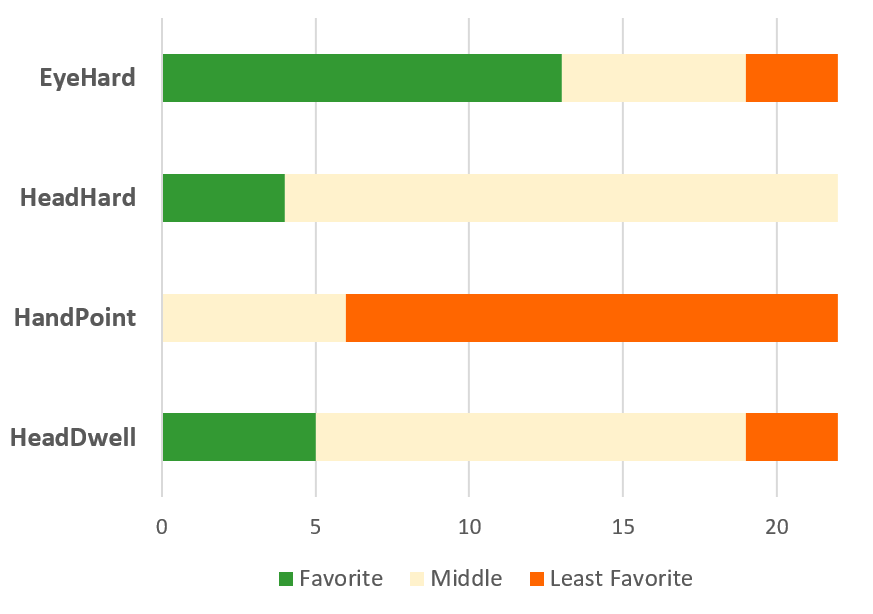}
    \Description{A bar chart showing user preferences. Green means favorite, yellow means middle and red means least favorite. EyeHard is over half colored in green, HeadHard and HeadDwell are mostly covered in yellow and HandPoint is around 2/3 covered in red.}
    \caption{User Preferences grouped by condition. Participants stated their most and least favored method, everything else got assigned the \textit{Middle} score.}
    \label{fig:USER_PREFERENCES_BARCHART}
\end{figure}

The results of the reported user preferences are presented in Figure \ref{fig:USER_PREFERENCES_BARCHART}. Among the evaluated methods, EyeHardware was the most preferred, with 59.1\% ($N = 13$) of participants favoring it. HeadHardware ($N = 4; 18.2\%$) and HeadDwell ($N = 5; 22.7\%$) were similarly preferred. Conversely, HandPoint was the least preferred option, with 72.7\% ($N = 16$) of participants selecting it as their least favorite and no participant preferring it.

\section{Discussion}\label{chapter:discussion}
In this section, we discuss and analyze the findings obtained from our user study and semi-structured interview, in relation to the research questions that were previously established. Statements of most participants were given in german and were translated to english for better understanding.

As mentioned in chapter \ref{chapter:participants_and_apparatus}, seven participants that required prescription glasses did not use them during the study and thus had at least some vision limitations. None of these participants mentioned any limitations due to impaired vision during the study and in the semi-structured interview. In addition, there was no significant effect of those participants on SUS ($t(86)=0.446, p=0.657$) and RTLX ($t(86)=-0.560, p=0.577$) scores.

\subsection{Influences on Speed and Error Rate} \label{chapter:discussion_speed_error}
Using the data recorded in our study software and presented in Tables \ref{tab:TIME_DESCRIPTIVES} and \ref{tab:SELECTIONS}, we address the research question \textbf{Q1:} \textit{Which selection method offers the highest selection speed and lowest error rate?} in the following paragraphs.

Overall, techniques utilizing a hardware button demonstrated the highest performance in terms of both selection time and time elapsed between hover and selection. Qualitative interviews conducted with study participants revealed that many described these techniques as "easy" and "fast" when asked to provide positive feedback. These findings are consistent with those reported in a previous study by Kyoto et al. \cite{kyto2018pinpointing}.

The mean select time for the HeadDwell technique with a one-second dwell time ($M = 2.45s$) was found to be 0.5 seconds slower compared to the mean select time of the HeadHard method ($M = 1.95s$). This difference is nearly identical to the mean delta time between select and hover for HeadDwell ($M = 0.993s$) and HeadHard ($M = 0.494s$), which is $0.499$. Therefore, the overall slower performance of HeadDwell can be attributed solely to the dwell time of one second, in contrast to the approximately 0.5 seconds it took to press the hardware button for HeadHard. It is worth noting that none of the participants found the chosen dwell time to be significantly too low or high. Participants were not aware of the Midas touch problem, but most stated that they were aware of accidental selections when using HeadDwell.

HandPoint was the slowest technique regarding selection time ($M = 2.76s$) and the second slowest technique regarding the delta time between select and hover ($M = 0.930s$), only being faster than HeadDwell. 
Several participants, especially those with little or no experience with AR interaction techniques, needed time to learn how to move their hands efficiently, react to instances where the system lost track of their hands, and perform the air tap gesture efficiently. Regarding speed, participants mostly expressed negative opinions on HandPoint such as "the technique is not that intuitive and I first needed some practice" and "this one is more complicated, I had to concentrate more on the selection".

Concerning error rate, head-based techniques had the highest hit-rate on marked POIs, as shown in Table \ref{tab:SELECTIONS}, with both HeadDwell and HeadHard being perceived as highly reliable. Positive feedback for HeadDwell included multiple participants noting the technique's precision, with comments such as "really precise when the car is not shaking or in curves" and "I had a high hit rate with this one". Similarly, positive feedback for HeadHard included comments such as "it was very clear what was happening" and also noting the precision when not influenced by g-forces.

The low percentage of selections that were missed for both techniques can be attributed to the high influence of car movements on the direction of the head gaze. Participants could not comfortably rest their head on the seats headrest due to the HMDs headstrap and therefore could not use the head rest to support and stabilize their movements. Most negative feedback regarding the head-based techniques was related to this issue. HeadDwell received negative comments such as "Sometimes I lost the dwell in curves", "It was difficult to hold the cursor on the element", "I had to steer my head against the g-forces in curves". Similarly, comments regarding HeadHard included "The heavy headset was greatly influenced by the car forces" and "I had to press the button multiple times at speedbumps or in sudden curves".

HeadDwell's low percentage of incorrect selections ($0.71\%$) is noteworthy given the Midas touch problem of implicit selection confirmation. The rather long dwell time of one second \cite{riegler2020gaze} and the participant's cautiousness could explain this finding. Some participants reported that the HeadDwell technique restricted their ability to freely look around and required more concentration to avoid unintentional selections with statements such as "I could not look around freely" and "I had to concentrate more to not select anything after selecting the marked one". Although statistical significance was not found for supporting this problem, HeadDwell was perceived as more mentally demanding than the hardware-based techniques, as demonstrated by higher mean scores on the RTLX-subscale mental and overall RTLX score.

EyeHard had the highest percentage of false selections (2.86\%) and the second highest percentage of missed selections (7.79\%). This could potentially stem from technical errors, either from the eye-tracking system, from the study software, or a combination of both. Some participants reported that a button press triggered two selections at once, indicating a technical issue. Some participants had problems, where the accuracy and eye-recognition were worse. Video recordings of the HMD-view in those cases showed the eye gaze pointer being offset a few degrees up or down. Another problem that was mentioned was called "flickering," where the gaze cursor jittered at the border of the POI hitbox, therefore triggering multiple hovers every few milliseconds and making selection more difficult as the user had to press the button at the right time. This error was also visible in some of the recordings of the headset view. To fix this error, a small cooldown window where the hover stays active after hovering off a POI could be implemented.

HandPoint had the highest percentage of missed selections overall, with 30\% of POIs not being selected. This technique faced the most technical limitations, despite the use of state-of-the-art hardware. Many participants struggled with performing the air tap gesture, and the system frequently lost track of their hands, making selection difficult. Participants also reported feeling restricted in their movements and not wanting to disturb the driver while gesturing around. In particular, selecting the rightmost POIs posed a challenge for HandPoint. Most participants attempted to select them with their right hand, resulting in hitting their hand against the door or needing to rotate their upper body or twist their arm. Some participants even resorted to using their left hand, which required additional mental effort. For example, one participant stated "First, I had to think about which hand to use before I could even think about the rest".

\subsection{Influences on Workload and Usability}
The research question \textbf{Q2:} \textit{Does the selection method have an impact on perceived workload and usability?} will be mainly answered with results stemming from the RTLX and SUS questionnaires, as described in Chapters \ref{chapter:Workload} and \ref{chapter:Usability}. The used selection method had significant influence on both perceived workload and usability. Inspection of the data in Figures \ref{fig:BOXPLOT_TLX_SCORE} and \ref{fig:BOXPLOT_SUS_SCORE} as well as pairwise comparisons revealed HandPoint to be the deviating method. 

HandPoint showed to have significantly higher mean scores in most subscales (performance, effort, frustration, and temporal demand) as well as in the overall scores for RTLX and SUS compared with the other methods. Lower perceived performance is in line with the measured performance shown in the high error and missing rates explained in Chapter \ref{chapter:discussion_speed_error} and listed in Table \ref{tab:SELECTIONS}. Higher effort is also in line with already described problems within HandPoint, like the difficulities participants had with the tracking and the confirmation gesture. Frustration was called out verbally by some participants who had problems with the technique. Higher temporal demand e.g. participants feeling time pressure is also reflected by the slower average selection time and DeltaSelectHover time discussed in the previous chapter. Those problems probably had the similar influence on the usability score.

The study found no statistically significant differences in the RTLX subscales and workload score for the pairings of Head\-Dwell, Head\-Hardware, and Eye\-Hardware. However, the mean values show, that Eye\-Hard had the lowest workload, followed by Head\-Hard and then Head\-Dwell. 
One possible explanation for this is the difference in neck strain between the techniques. Eye gaze requires minimal head movement and participants can quickly glance at the POI and then reset to a more comfortable position after a quick button press. Additionally, the FoV for the Varjo XR-3 is around 115° \cite{VarjoXR3}, which means for eye gaze, that the POI just needs to be visible in this window. In contrast, head gaze requires the POI to be at the center of the view which requires more head movement and may cause neck strain. Participants reported that the movements for the head pointing techniques felt unnatural, and they experienced difficulty keeping their head steady and discomfort in the neck when viewing outermost POIs. This was reinforced by the HeadDwell condition where the head had to remain in those more unnatural positions for one second. These findings align with those of Kyoto et al. \cite{kyto2018pinpointing}, where participants found eye pointing easy and fast, but reported difficulties in keeping their head steady and discomfort in the neck.

Regarding usability, HandPoint was the worst rated method out of the four. DSCF pairwise comparisons showed significant differences between HandPoint and each other technique. The other techniques had similar mean scores. For easier interpretation, Bangor et al. \cite{bangor2009sus} assigned adjectives to ranges of SUS scores. According to his ratings, the usability of the HandPoint technique could be described as \textit{ok} ($Mean = 58.0, "ok" > 50.9$), HeadDwell as \textit{good} ($Mean = 83.8, "good" > 71.4$), EyeHard as \textit{excellent} ($Mean = 85.6, "excellent" > 85.5$), and HeadHard also as \textit{excellent} ($Mean = 87.6, "excellent" > 85.5$).

\subsection{Influences on User Preference}
Results of the semi-structured interview and participant's statements during the study were used to answer the research question \textbf{Q3:} \textit{Which selection methods are the most and least preferred by participants?}. User preferences are described in Chapter \ref{chap:user_preference} and shown in Figure \ref{fig:USER_PREFERENCES_BARCHART}.

EyeHardware was clearly the preferred method of most participants, with 59.1\% ($N = 13$) favoring it. At the same time, it was intuitive and fast ("It can't get any simpler than that") while still coming off as striking ("It was kind of like magic"). The participants that picked EyeHard as their least favorite in our study all had technical problems with the tracking system. Our results are in line with Blattgerste et al. \cite{blattgerste2018advantages}, where eye gaze was preferred over head gaze. However, these results are contrary to the study Kyoto et al.~\cite{kyto2018pinpointing} conducted, where head interactions were slightly preferred over eye gaze. For them, eye pointing was also described as easy and fast, but also apparantly not accurate enough. This difference could stem from the limited FoV of the HoloLens of 30x17° compared with the 115° of the Varjo XR-3 \cite{VarjoXR3}, as advantages of eye gaze over head gaze are dependent on FoV \cite{blattgerste2018advantages}.

HeadHardware ($N = 4; 18.2\%$) and HeadDwell ($N = 5; 22.7\%$) were similarly preferred. Apart from the neck strain some participants felt after using these techniques non-stop for seven minutes and the influence of g-forces on the head movements, there were not many negative comments about the techniques. They performed precise and reliably but were not the fastest and not the most exciting. Noteworthy, HeadHardware was the only method that was not selected as the least favorite by any of the participants.

Conversely, HandPoint was the least preferred option with 72.7\% ($N = 16$) selecting it as their least favored. Notably, hand pointing was not chosen as a favorite method by any of the participants. While our study is not perfectly comparable with the study of Kyoto et al. \cite{kyto2018pinpointing}, some connections can be drawn nevertheless, as their combination of head pointing and eye pointing with gesture confirmation were also both among the least favored methods.

\subsection{Limitations}
As mentioned in chapter \ref{chapter:participants_and_apparatus}, one participant's data was excluded due to almost immediate symptoms of motion sickness and the following termination of the study run. We plan to employ the Motion Sickness Susceptibility Questionnaire \cite{golding1998motion} in future studies to mitigate the risk of motion sickness in participants.

Although we performed initial eye-tracking calibration for all participants, we cannot exclude that errors might have occurred within this process. They could originate from user error, as well as from issues within the technical system. Within our study, this resulted in an offset of the eye-tracking cursor, impacting selection precision for some participants. Furthermore, such errors could also have resulted in flawed video see-through quality. 

The employed 6-DoF tracking system did not perform without flaws, especially within the first minute of driving. Here, errors concerning the alignment of the car-fixed coordinate system resulted in displayed POIs drifting away or changing position and rotation in relation to the car. This caused irritation across participants, but also impacted the active round of the task, as targets were moving and therefore harder to select. After a few seconds, the issue resolved itself, with POIs returning to their initial position, allowing the study to proceed as usual. Additionally, a minor misconfiguration of the tracking system's coordinates resulted in a slight shift of POI alignment to the right, making POIs on the right side more challenging to reach than those on the left. This offset was consistent accross all participants.

Furthermore, we attempted to synchronize the length of the driven course with the required duration of a complete trial. However, this was not always successful across all participants, and most of them ended up performing the last few rounds in standstill instead of the intended investigation in a moving vehicle. As this behavior is also realistic for real world scenarios, such as waiting at a traffic light, and was consistent across participants, we do not anticipate any significant impact on the results.

Finally, the built-in hand tracking occasionally failed to recognize participants' hands, which may have been due to dynamic lighting in an outdoor environment, where bright sunlight can reduce the performance of hand tracking. Furthermore, the air tap gesture for selection was occasionally not detected by the system, resulting in longer selection times, more missed selections and higher frustration. Further studies testing the LeapMotion for in-car interaction may give more insight into these problems.

\section{Conclusion and Future Work}
In conclusion, we explored four different selection techniques for digital objects outside of moving vehicles and identified which techniques are best suited for use with AR HMDs. Our findings bridge the gap between conventional AR interaction research in stationary environments and the practical implementation of AR technology in the automotive space. Additionally, we emphasize the significance of testing HMD-based AR systems in real-world scenarios, as it can yield valuable insights necessary for the successful deployment of such systems for consumers in the future.

Our investigation included: head pointing with dwell time, head pointing with a hardware button, eye gaze with a hardware button, and hand pointing with gesture confirmation. Our user study conducted under real-world conditions demonstrated that eye gaze combined with a hardware button was the fastest and preferred technique with the lowest perceived workload. While head-based techniques were not as much preferred as eye gaze, they had a lower error rate, and similarly low perceived workload values and similarly high usability scores. In contrast, hand pointing with gesture confirmation was the least favored technique across all tested categories. Overall, these findings provide valuable insights into the selection of digital objects in AR outside of moving vehicles and contribute to the field of in-car AR-based HCI research.

Following our findings, we want to improve the tested techniques, for example by exploring the combination of multiple techniques to employ a multimodal approach and by integrating them more tightly into the car's ecosystem. Additionally, we plan to investigate interaction with location-based data, explore other environments like highways, map the track characteristics to the data, and explore more advanced interaction techniques, such as creating or manipulating digital objects in AR.


\bibliographystyle{ACM-Reference-Format}
\bibliography{bibliography}



\end{document}